\documentclass[showpacs,twocolumn]{revtex4}

\usepackage{graphicx,amsmath,psfrag,color} 
\usepackage{latexsym}
\usepackage{exscale}
\usepackage{epstopdf}
\usepackage{amssymb,mathrsfs}

\newcommand{\bs}[1]{\boldsymbol{#1}}
\newcommand{\vac}{\left|\,0\,\right\rangle}
\newcommand{\ket}[1]{\left|#1\right\rangle}

\newcommand{\up}{\uparrow}
\newcommand{\dw}{\downarrow}
\newcommand{\nn}{\nonumber}
\definecolor{hellgrau}{gray}{.5}
\definecolor{hhellgrau}{gray}{.9}
\def\ie{{\it i.e.},\ }
\def\eg{{\it e.g.}\ }
\def\ea{{\it et al.}}

\allowdisplaybreaks[4]
\begin{document}

\title{Spin $\bs{3/2}$ dimer model}
\author{Stephan Rachel} 
 \affiliation{Institut f\"ur Theorie der Kondensierten Materie, 
   Universit\"at Karlsruhe, Postfach 6980, 76128 Karlsruhe, Germany}
\pagestyle{plain}
\begin{abstract}
  We present a parent Hamiltonian for weakly dimerized valence bond
  solid states for arbitrary half--integral $S$. While the model
  reduces for $S=1/2$ to the Majumdar--Ghosh Hamiltonian we discuss
  this model and its properties for $S=3/2$. Its degenerate ground
  state is the most popular toy model state for discussing
  dimerization in spin 3/2 chains. In particular, it describes the
  impurity induced dimer phase in Cr8Ni as proposed recently.  We
  point out that the explicit construction of the Hamiltonian and its
  main features apply to arbitrary half--integral spin $S$.
\end{abstract}

\maketitle

\section{Introduction}
Valence bond solid (VBS) states played an important role for the
understanding of quantum spin chains. Originally introduced as exact
ground states of the integral spin $S$ Affleck--Kennedy--Lieb--Tasaki
(AKLT) models~\cite{affleck-87prl799,affleck-88cmp477}, VBS states
meanwhile refer to all spin $S$ states consisting of 
nearest--neighbor singlet bonds. Amongst them, the spin 1/2
Majumdar--Ghosh (MG)
states~\cite{majumdar-69jmp1388,majumdar-69jmp1399} are the most
important examples as their parent Hamiltonian corresponds to a
certain point in the parameter space of the spin 1/2 $J_1$--$J_2$
model with antiferromagnetic $J_i$. As such, the MG model links
dimerized VBS states and dimerized spin chains
where the frustration is forced by next--nearest neighbor
interactions. For $S>1/2$, however, this exact link is not given
anymore. The VBS states seem not to belong to the phase diagrams of
the frustrated Heisenberg chains.  Nonetheless, VBS states are often
used as toy model states in order to discuss frustration in quantum
spin chains, see for the spin 3/2 case for example Refs.
\cite{affleck89jpcm3047,guo-90prb9592,niggemann-96zpb289,
  oshikawa-97prl1984,roth-98prb9264,nakamura-02prl077204,
  almeida-08prb094415,almeida-09prb115141,refael-02prb060402} and
references therein. Parent Hamiltonians, however, are only known for
the Majumdar--Ghosh states~\cite{majumdar-69jmp1388}, for the
translational invariant AKLT states~\cite{affleck-87prl799}, and for
the spin 1 dimer states~\cite{affleck89jpcm3047}. For spin 3/2 and
larger $S$, no exact and translational invariant Hamiltonians are
known having dimerized VBS states as their unique ground states.
Most recently, VBS states have attracted renewed
interest. They have been generalized to special unitary SU($n$)
symmetry~\cite{chen-05prb214428,greiter-07prb060401,greiter-07prb188841,arovas-08prb104404},
to special orthogonal symmetry
SO($n$)~\cite{tu-08jpa415201,tu-08prb094404}, and to symplectic
symmetry SP($n$)~\cite{schuricht-08prb014430}. A supersymmetric
extension of VBS states has been proposed
recently~\cite{arovas-09arXiv:0901.1498}.

\begin{figure}[t]
\centering
\includegraphics[scale=1.0]{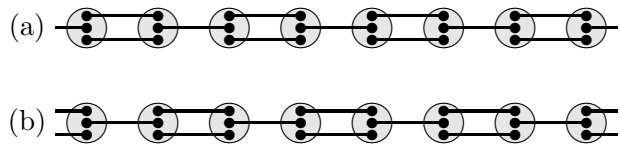}
\caption{Illustration of the spin $3/2$ dimer states. (a) and (b)
  are the two--fold degenerate partially dimerized VBS states. Each
  state can be seen as the covering of the AKLT state with one of the
  Majumdar--Ghosh states. The large grey--shaded circles are the lattice
  sites, while the small black dots denote a spin 1/2. The spins 1/2
  on a lattice site will be projected onto the totally symmetric
  subspace, \ie $S=3/2$. The lines between two spins 1/2 denote a
  singlet bond.}
\label{fig:S=3/2-dimer-states}
\end{figure}

In this Letter, we present a parent Hamiltonian for the partially or
weakly dimerized VBS states 
(see Fig.\,\ref{fig:S=3/2-dimer-states} for the illustration of the
spin 3/2 case).  For spin 1/2, this Hamiltonian reduces to the MG
Hamiltonian. The particular importance of the spin 3/2 Hamiltonian is
given by its possible realization in the impurity induced dimer
phase in Cr8Ni as proposed recently~\cite{almeida-09prb115141}.
This molecular nanomagnet comprises eight chromium(III) ions with spin
3/2 and one nickel(II) ion with spin 1. The ground state of this Cr8Ni
ring was shown to be well understood in the VBS
picture~\cite{almeida-09prb115141} by means of numerical
simulations.
Likewise, the Hamiltonian can be implemented within DMRG, as
only standard spin interaction terms are present. This will allow for extensive
numerical studies complementary to known results for the spin 3/2
$J_1$--$J_2$ model with antiferromagnetic $J_i$~\cite{roth-98prb9264}.

The Letter is organized as follows: we start with a brief discussion
of general properties of VBS states. Then we introduce the $S=3/2$
dimer Hamiltonian which has the weakly dimerized VBS states shown in
Fig.\,\ref{fig:S=3/2-dimer-states} as ground states and show that
these are the unique ground states. We show that the lowest lying
excitations are fractionally quantized spin 1/2 spinons.  Then we
discuss the connection to the dimerized phase in the frustrated spin
3/2 Heisenberg model and static spin--spin correlations. Finally, we
show that the explicit construction of the Hamiltonian as well as all
qualitative results generalize to the case of arbitrary half--integral
spin $S$.

\section{VBS States}
\begin{figure}[t]
\centering
\includegraphics[scale=1.0]{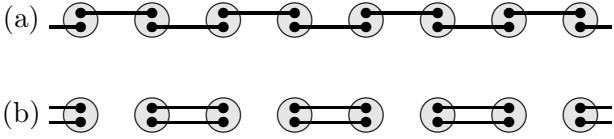}
\caption{The $S=1$ VBS states: (a) is the translational invariant AKLT
  state and (b) is one the of two--fold degenerate dimer states. The
  second dimer state can be obtained by shifting the state (b) by one
  lattice spacing to the right.}
\label{fig:S=1-states}
\end{figure}

In 1987, Affleck, Kennedy, Lieb and Tasaki (AKLT) introduced a spin 1
Hamiltonian which was constructed as a sum over a local projection
operator. Its unique ground state can be understood as follows: on
each lattice site are two spins 1/2 symmetrically coupled into an
effective spin 1, while each of the spins 1/2 is antisymmetrically
coupled to a spin 1/2 on its neighboring lattice sites, either to the
right or the left. For an illustration see
Fig.\,\ref{fig:S=1-states}~(a). This state was called a valence bond
solid. We wish to understand as a VBS all states consisting of local
singlet bonds. In particular, we call the MG states and the spin
1 dimer states (see Fig.\,\ref{fig:S=1-states}~(b)) also VBS states
even though they break translational invariance spontaneously.

One of the main features of VBS states is that we are able to write
down their wave functions explicitly as easily as we illustrate them.
Schwinger
bosons~\cite{schwinger65,auerbach94,arovas-08arXiv:0809.4836}
constitute a convenient way to formulate spin $S$ representations.
Arovas \ea~\cite{arovas-88prl531} applied for the first time Schwinger
bosons to the AKLT model and to VBS states in general. The spin $S$
operators
\begin{equation}
  S^+=a^{\dagger}b,~~~S^-=b^{\dagger}a,~~~S^z=\frac{1}{2}\left(a^{\dagger}a
-b^{\dagger}b\right)
\end{equation}
are expressed by bosonic creation and annihilation operators which
obey the standard commutation relations
$[a,a^{\dagger}]=[b,b^{\dagger}]=1$ while all other commutators
vanish. The spin quantum number $S$ is given by $2S=a^{\dagger}a
+b^{\dagger}b$ and the general spin $S$ state is defined as
\begin{equation}
  \ket{S,s_z}=\frac{\left(a^{\dagger}\right)^{S+s_z}}{\sqrt{(S+s_z)!}}
\frac{\left(b^{\dagger}\right)^{S-s_z}}{\sqrt{(S-s_z)!}}\vac.
\end{equation}
The spin 1/2 states are thus given by
$c_{i\up}^{\dagger}\vac=a_i^{\dagger}\vac$ and
$c_{i\dw}^{\dagger}\vac=b_i^{\dagger}\vac$, respectively.  The
difference between the fermionic operators and Schwinger bosons shows
up only when two or more creation operators act on the same site.
While fermion operators create a singlet configuration
$c_{i\up}^{\dagger}c_{i\dw}^{\dagger}\vac$, Schwinger bosons create a
totally symmetric representation.

The elegance of Schwinger bosons come to light when defining the
general spin $S$ VBS state:
\begin{equation}
  \label{gs-general}
\ket{\psi_{(m,n)}}~=~\prod_i ~\left(\mathcal{B}_{2i}\right)^{\,m}
\left(\mathcal{B}_{2i-1}\right)^{\,n}\vac,
\end{equation}
where $\mathcal{B}_{i}=
a^{\dagger}_{i}b^{\dagger}_{i+1}-b^{\dagger}_{i}a^{\dagger}_{i+1}$
creates on sites $i$ and $i+1$ a singlet bond. We recover for $(m=1,n=0)$
and $(0,1)$, respectively, the MG states. The $S=1$ AKLT state can
thus be written as $\ket{\psi_{(1,1)}}$ and the spin 1 dimer state
(Fig.\,\ref{fig:S=1-states}~(b)) as $\ket{\psi_{(2,0)}}$. Obviously,
$m+n=2S$ and there are always $2S+1$ VBS states. For half--integer
$S$, there are always $S+\frac{1}{2}$ pairs of dimerized states with
different strengths of dimerization. For integer $S$, there are $S$
pairs of dimerized states and one additional AKLT state
$\ket{\psi_{(S,S)}}$. Note that there is always a pair of completely
dimerized VBS states regardless $S$.  We stress again that the
Schwinger bosons ensures the symmetric projection on every lattice
site.

\begin{figure}[t]
\centering
\includegraphics[scale=1.0]{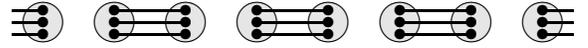}
\caption{Illustration of one of the two--fold degenerate completely
  dimerized $S=3/2$ VBS states, say $\ket{\psi_{(0,3)}}$. By 
  shifting of one lattice spacing to
  the right we obtain the second dimer state $\ket{\psi_{(3,0)}}$.}
\label{fig:S=3/2-fullydimerized}
\end{figure}

In this Letter, we are mainly interested in $S=3/2$ VBS states. There
are four of them: completely dimerized states $\ket{\psi_{(3,0)}}$ and
$\ket{\psi_{(0,3)}}$ (the latter is shown in
Fig.\,\ref{fig:S=3/2-fullydimerized}) and weakly dimerized states
$\ket{\psi_{(2,1)}}$ and $\ket{\psi_{(1,2)}}$ (both are shown in
Fig.\,\ref{fig:S=3/2-dimer-states}). For the weakly dimerized VBS
states we present in the next paragraph a parent Hamiltonian.
We wish to mention that one can easily define a spin-$S$ Hamiltonian
in terms of local projection operators annihilating all $2S+1$ VBS
states.  It involves three--site interactions and reduces for $S=1/2$
to the well--known MG Hamiltonian.  For $S>1/2$, however, this is not
a parent Hamiltonian anymore as it has all spin $S$ VBS states as
groundstates.  The Hamiltonian is given by
\begin{equation}\label{H1-S}
\mathcal{H}_{1}(S)=\sum_i 
\sum_{\mu=S+1}^{3S} A_{\mu}P^{(\bs{\mu})}_{i,i+1,i+2}\ .
\end{equation}

\section{Spin $\bs{3/2}$ Dimer Hamiltonian}
Now we present the Hamiltonian which uniquely singles out the weakly
dimerized VBS states $\ket{\psi_{(2,1)}}$ and
$\ket{\psi_{(1,2)}}$ as shown in Fig.\,\ref{fig:S=3/2-dimer-states}. This
Hamiltonian is a sum of projection operators
$P^{(\bs{S})}_{i,i+1,\ldots,i+\mu}$ which project onto the spin
$S$ subspace on $\mu+1$ neighboring sites.
\begin{equation}
  \nn 
\begin{split}
  \mathcal{H}^{\rm dimer} = \sum_i \Big(&~
A_{\frac{5}{2}}P^{(\bs{\frac{5}{2}})}_{i,i+1,i+2}
+ A_{\frac{7}{2}}P^{\bs{(\frac{7}{2}})}_{i,i+1,i+2} \Big. \\[7pt]
\Big. &+ A_{\frac{9}{2}} P^{(\bs{\frac{9}{2}})}_{i,i+1,i+2}
+ \beta \, P^{(\bs{3})}_{i,i+1} ~\Big)
\end{split}
\end{equation}
Remarkably, the last projection operator acts on two neighboring sites
while the other operators act on three neighboring sites. We
decompose, hence, the Hamiltonian into two parts, $\mathcal{H}_1$
acting on three neighboring sites and $\mathcal{H}_2$ acting on
neighboring sites. We are able to express
$\mathcal{H}_1=\mathcal{H}_1(\frac{3}{2})$ by means of spin operators
as follows,
\begin{equation}
  \label{ham:3site-projection}
\begin{split}
  \mathcal{H}_1 = \sum_i & \left( 
\left(\bs{S}_i + \bs{S}_{i+1} + \bs{S}_{i+2} \right)^2-\frac{3}{4} \right)\\
&\times
\left( \left(\bs{S}_i + \bs{S}_{i+1} + \bs{S}_{i+2} \right)^2-\frac{15}{4} 
\right).
\end{split}
\end{equation}
The second part, $\mathcal{H}_2$, corresponds to the nearest neighbor
bilinear--biquadratic--bicubic Heisenberg model with certain parameters,
\begin{equation}
  \label{ham:2site-projection}
\begin{split}
  &\mathcal{H}_2 = \sum_i P^{(\bs{3})}_{i,i+1} =\\
  &\sum_i \Bigg[
\bs{S}_i\bs{S}_{i+1} + \frac{116}{243}\left(\bs{S}_i\bs{S}_{i+1}\right)^2
+\frac{16}{243}\left(\bs{S}_i\bs{S}_{i+1}\right)^3 + \frac{55}{108} \Bigg].
\end{split}
\end{equation}
Finally, the parent Hamiltonian  can be written for $\beta>0$ as
\begin{equation}\label{ham:H1+betaH2}
  \mathcal{H}^{\rm dimer}=\mathcal{H}_1+\beta\mathcal{H}_2.
\end{equation}
Note that this Hamiltonian can be implemented both within exact
diagonalization (ED) and DMRG. Terms like $(\bs{S}_i\bs{S}_j)^2$ and
$(\bs{S}_i\bs{S}_j)^3$ are standard terms. More complicated terms like
$(\bs{S}_i\bs{S}_j)(\bs{S}_i\bs{S}_k)$ or
$(\bs{S}_i\bs{S}_j)(\bs{S}_j\bs{S}_k)$ are also straight forward to
implement. The reader may notice that DMRG has been successfully
applied by several
authors~\cite{qin-95prb12844,hallberg-96prl4955,carlon-04prb144416,fath-06prb214447,heidrich-meisner-07prb064413}
to investigate antiferromagnetic spin 3/2 chains.

In the following we will prove that \eqref{ham:H1+betaH2} has a
two--fold degenerate zero-energy ground state,
\begin{equation}
  \mathcal{H}^{\rm dimer}\ket{\psi_{(2,1)}}=
  \mathcal{H}^{\rm dimer}\ket{\psi_{(1,2)}}=0.
\end{equation}
The proof is exceedingly simple: first we will verify that
$\mathcal{H}_1$ has the four $S=3/2$ VBS states as unique zero--energy
groundstates, second we will show that $\mathcal{H}_2$ exhibits a huge
(actually an infinite) ground state degeneracy (also with zero energy)
which contains the weakly dimerized VBS states but not the completely
dimerized VBS states.  Consequently, the additive term
$\beta\mathcal{H}_2$ in \eqref{ham:H1+betaH2} does not affect the
weakly dimerized VBS state while the fully dimerized state will be
lifted to higher energy. This lifting is controlled by the parameter
$\beta>0$ and can be tuned continuously.

The auxiliary Hamiltonian $\mathcal{H}_{1}$ has the four $S=3/2$ VBS
states as exact zero--energy ground states,
$\mathcal{H}_{1}\ket{\psi_{(3,0)}}=\mathcal{H}_{1}\ket{\psi_{(2,1)}}=
\mathcal{H}_{1}\ket{\psi_{(1,2)}}=\mathcal{H}_{1}\ket{\psi_{(0,3)}}=0$.
\begin{figure}[t!]
  \centering
  \includegraphics[scale=1.0]{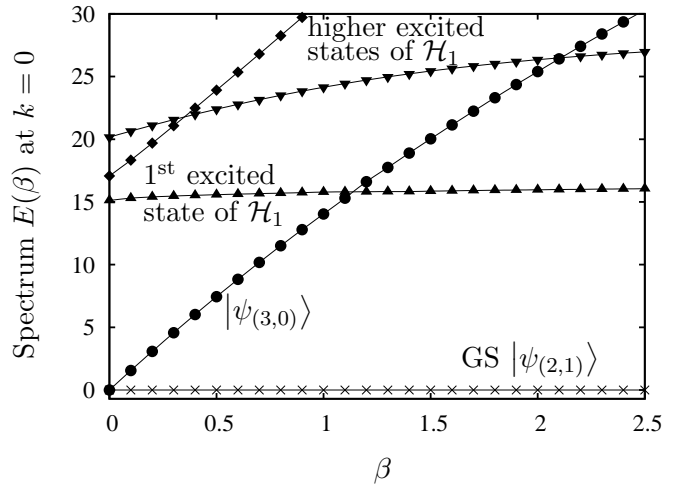}
  \caption{Lowest lying energies of the Hamiltonian
    \eqref{ham:H1+betaH2} as a function of $\beta$ at the
    $\Gamma$--point (momentum $k=0$) in the Brillouin zone. The data
    was obtained on a chain with $N=12$ lattice sites and periodic
    boundary conditions. As the $\Gamma$--point is considered, there
    are only two of the four groundstates for $\beta=0$. The other
    groundstates (\,$\ket{\psi_{(1,2)}}$ and $\ket{\psi_{(0,3)}}$\,)
    are located at $k=\pi/a$ in the Brillouin zone. For increasing
    $\beta>0$ the fully dimerized VBS state $\ket{\psi_{(3,0)}}$ is
    increased in energy.  Around $\beta\approx 1.1$ it crosses with
    the first excited state of $\mathcal{H}_1$ which is in some sense
    also the first excited state of $\mathcal{H}^{\rm dimer}$. This
    crossing is expected, but the value $\beta\approx 1.1$ depends on
    the system size and will decrease for larger $N$.}
  \label{fig:ham-beta}
\end{figure}

This can be seen as follows: Consider three neighboring sites of one
of the four VBS states (no matter which of the VBS states, completely
or partially dimerized ones): there are always three singlet bonds and
three individual spin 1/2 representations,
$\bs{0}\otimes\bs{0}\otimes\bs{0}\otimes\bs{\frac{1}{2}}
\otimes\bs{\frac{1}{2}}\otimes\bs{\frac{1}{2}}=2\cdot\bs{\frac{1}{2}}
\oplus\bs{\frac{3}{2}}$.  In general, the tensor decomposition of
three spin 3/2 is given by
$\bs{\frac{3}{2}}\otimes\bs{\frac{3}{2}}\otimes\bs{\frac{3}{2}}=
2\cdot\bs{\frac{1}{2}}\oplus 4\cdot\bs{\frac{3}{2}}\oplus
3\cdot\bs{\frac{5}{2}}\oplus 2\cdot \bs{\frac{7}{2}}\oplus
\bs{\frac{9}{2}}$. The projection onto the subspace with
representation $S=5/2$, $S=7/2$, and $S=9/2$ singles out the four VBS
states as ground states. We can express the local projection operator
in terms of spin operators as done in equation
\eqref{ham:3site-projection}. Now we consider the nearest neighbor
projection operator $P^{(\bs{3})}_{i,i+1}$. The Hamiltonian
$\mathcal{H}_{2}=\sum_i P^{(\bs{3})}_{i,i+1}$ has an infinite ground
state degeneracy. This can be seen as every state is a ground state
where two spins 1/2 per site are in the ``AKLT configuration'' and the
third spin 1/2 is ``free''. The partially dimerized states obviously
belong to this set of ground states while the completely dimerized
states are not ground states of $\mathcal{H}_2$.  Consequently, the
term $\beta\mathcal{H}_2$ in \eqref{ham:H1+betaH2} lifts the
completely dimerized states continuously to higher energy as $\beta$
is increased. The partially dimerized states remain as unique ground
states. This scenario is shown in Fig.\,\ref{fig:ham-beta} for a chain
with $N=12$ sites and periodic boundary conditions where the data was
obtained by means of exact diagonalization.

For open boundary conditions the situation changes: the dimer states
are no global singlets anymore, edge states emerge, and the dimer
states are fourfold (singlet and triplet) and ninefold (singlet,
triplet, and quintuplet) degenerate. When cutting the dimer states,
either one individual spin 1/2 or two symmetrically coupled spins 1/2
remain which are not coupled in local singlet bonds (the two different
cuts correspond to the two ground states in case of periodic boundary
conditions). The {\it dangling} spins at the edges cause the
degeneracy of the dimer ground states. For the first situation ($N$
even) the two edge spins couple into a singlet and a triplet,
$\bs{\frac{1}{2}}\otimes\bs{\frac{1}{2}}=\bs{0}\oplus\bs{1}$, yielding
a four--fold degenerate dimer state. For the second situation ($N$
even), the two spins 1/2 at each edge are symmetrically coupled into a
triplet,
$\mathcal{S}\{\bs{\frac{1}{2}}\otimes\bs{\frac{1}{2}}\}=\bs{1}$. Hence,
the two edge spins carry spin 1 and will be coupled into singlet,
triplet, and quintuplet,
$\bs{1}\otimes\bs{1}=\bs{0}\oplus\bs{1}\oplus\bs{2}$, yielding a
ninefold degenerate ground state. This manifold degeneracy of the
ground state in case of open boundary conditions can be nicely
observed numerically, when looking at the different $S^z$--subspaces.

\section{Fractionally quantized excitations}
\begin{figure}[t]
  \centering
  \includegraphics[scale=1.0]{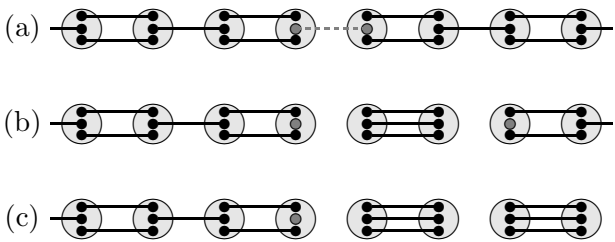}
\caption{(a) The only possibility to excite the system being in one of
  the VBS ground states is to break a singlet bond (indicated by the
  dashed line). (b) The resulting spinons (the small dark--grey dots)
  are deconfined and can freely move through the chain as they
  interpolate between different ground states.  In fact, the left
  spinon in the cartoon constitutes a domain wall between the ground
  states $\ket{\psi_{(2,1)}}$ and $\ket{\psi_{(3,0)}}$ while the right
  spinon constitutes a domain wall between $\ket{\psi_{(3,0)}}$ and
  $\ket{\psi_{(2,1)}}$. (c) For clarity we have fixed the left spinon.
  From the sequence (a), (b), and (c) we note that the spinons can
  occupy only half the number of sites (see the explanation in the
  text). The reader should notice that the ``states'' shown in (a),
  (b), and (c) are not exact eigenstates of $\mathcal{H}_1$ but the
  cartoons provide a good intuitive picture of the low--lying
  excitations.}
\label{fig:exc-H1}
\end{figure}

As already discussed, we are able to define the model Hamiltonian
$\mathcal{H}_1(S)$ (see Eq.\,\eqref{H1-S}) acting on three neighboring
sites for arbitrary spin $S$ exhibiting a $(2S+1)$--fold degenerate
ground state, \ie the $2S+1$ possible spin $S$ VBS
states~\cite{nakamura-02prl077204,sen-07prb104411}. For $S=1/2$, this
model corresponds to the MG Hamiltonian, for $S=1$ to the Hamiltonian
proposed by Sen and Surendran~\cite{sen-07prb104411}, and for $S=3/2$
to $\mathcal{H}_1$ of equation \eqref{ham:3site-projection}. In all of
these Hamiltonians, the lowest lying excitation will be a pair of
deconfined $S=1/2$ spinons. This is easy to understand as the spinons
emerge in this models as domain walls interpolating between different
ground states.  As the Hilbert space of spin $S$ antiferromagnets is
spanned by spin flips which carry spin 1, the spin 1/2 excitation has
to be seen as a fractionally quantized excitation. We illustrate this
situation for the Hamiltonian $\mathcal{H}_1$ in
Fig.\,\ref{fig:exc-H1}.
In the cartoon Fig.\,\ref{fig:exc-H1}, we notice another interesting
point: even though the spinons are free they can only occupy every second
lattice site. If there are only a few spinons in a chain the
number of orbitals available to them is roughly half the number of
sites reflecting their fractionally quantized character.

\begin{figure}[t]
  \centering
  \includegraphics[scale=1.0]{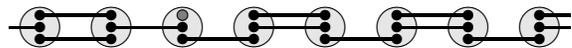}
\caption{The spinon excitation (the small dark--grey dot) interpolates
  between the states $\ket{\psi_{(2,1)}}$ and $\ket{\psi_{(1,2)}}$.
  The state on the right hand side of the spinon excitations is
  identical to the state in Fig.\,\ref{fig:S=3/2-dimer-states}\,b as
  explained in the text.}
\label{fig:exc-Hdimer}
\end{figure}

Now the question remains whether this picture of free spinon
excitations holds for the Hamiltonian $\mathcal{H}^{\rm dimer}$
considered in this paper.  As we mentioned above, relevant in our
cartoons are only the number of singlet bonds between neighboring
sites while the order of singlet bonds is irrelevant. Hence, a VBS
state is identified uniquely by $m$ and $n$.
The cartoon in Fig.\,\ref{fig:exc-Hdimer} suggests that elementary
excitations of \eqref{ham:H1+betaH2} are also spin 1/2 spinons.  This
is not unexpected as the Hamiltonian \eqref{ham:H1+betaH2} is the
Hamiltonian $\mathcal{H}_1$ plus an additional term $\mathcal{H}_2$
which lifts up the fully dimerized states out of the ground state
manifold. Altogether, the whole consideration of deconfined spinon
excitations in the spin 3/2 dimer model is consistent with the picture
that the origin of the Haldane gap~\cite{haldane83pla464} is a
confinement force between
spinons~\cite{greiter02jltp1029,greiter-07prb188841}. The generic
$S=3/2$ spin chain, the nearest neighbor Heisenberg model, is known
not to exhibit a Haldane gap and, hence, the spinons cannot be
confined. While the generic spin 3/2 chain has a gapless excitation
spectrum, the Hamiltonians $\mathcal{H}^{\rm dimer}$ and
$\mathcal{H}_1$ as well as all spin $S$ Hamiltonians mentioned at the
beginning of this section exhibit a gapped excitation spectrum as it
costs a finite energy to break a singlet bond. This energy cost
survives in the thermodynamic limit but it is not Haldane's gap.


\section{Discussion}
The Hamiltonian \eqref{ham:H1+betaH2} has the weakly dimerized states
$\ket{\psi_{(2,1)}}$ and $\ket{\psi_{(1,2)}}$, respectively, as exact
ground states. Dimerization is usually caused by frustration. In one
spatial dimension this is realized in a next--nearest neighbor
Heisenberg model, $\mathcal{H}=\sum_i \bs{S}_i\bs{S}_{i+1} + \alpha
\bs{S}_i\bs{S}_{i+2}$, when the next nearest neighbor coupling
$\alpha$ exceeds a critical value $\alpha_c$.

The spin 1/2 Heisenberg model is characterized by a critical phase for
$\alpha<\alpha_c=0.2411$~\cite{okamoto-92pla433,eggert96prbR9612,
  jullien-83baps344}.  The system undergoes a second--order phase
transition to a dimerized phase for $\alpha>\alpha_c$. For
$\alpha_{\rm MG}$, the Hamiltonian corresponds to the Majumdar--Ghosh
model exhibiting the exact dimer states $\ket{\psi_{(1,0)}}$ and
$\ket{\psi_{(0,1)}}$ as ground states.  Notice, that the dimerization
becomes exact in the thermodynamic limit only. That is, the ground
state becomes perfectly degenerate due to the broken translation
symmetry. One ground state has total momentum $k=0$ while the other
has $k=\pi$. Except for $\alpha_{\rm MG}$, the ground state degeneracy
is not intact in the finite system.  Nevertheless, ground states in
the dimer phase are known to exhibit a good overlap with the MG ground
states.

For the spin 3/2 Heisenberg model with next nearest neighbor
interaction $\alpha$, Roth and Schollw\"ock found the phase transition
to the dimerized phase at $\alpha_{c,3/2}\approx
0.29$~\cite{roth-98prb9264}. They further computed the maximum
dimerization for $\alpha_{\rm max}=0.415$.  We expect that the ground
states in the spin 3/2 dimer phase correspond to the weakly dimerized
states rather than the completely dimerized states, \ie our
Hamiltonian \eqref{ham:H1+betaH2} can be seen as the $S=3/2$--analogue
of the MG model. We can further conjecture that the states
$\ket{\psi_{(2,1)}}$ and $\ket{\psi_{(1,2)}}$ will have an excellent
overlap with the ground states in the dimer phase investigated by Roth
and Schollw\"ock for sufficiently large system lengths. Such system
lengths are not computable within ED.  Nevertheless, in principle the
Hamiltonian \eqref{ham:H1+betaH2} is implementable within DMRG as
pointed out above and the calculation of overlaps might be feasible.
Furthermore, it might be interesting to compute the static spin--spin
correlations within DMRG. Correlations are known to decay quite fast
in dimer phases, for the MG states, however, the correlations decay
abruptly, \ie $\langle \bs{S}_i\bs{S}_{i+2}\rangle=0$. Hence, the
question remains for the spin 3/2 dimer model whether the spin--spin
correlations decay fast, say exponentially, or abruptly. 

\section{Generalization to arbitrary half--integer $\bs{S}$}
We are able to formulate the generalized version of the Hamiltonian
\eqref{ham:H1+betaH2} to arbitrary half--integer $S$. It involves
interactions between three neighboring sites and is formulated in
terms of projection operators as follows:
\begin{equation}
  \label{ham-general}
  \mathcal{H}^{S}=\sum_i \left( 
\sum_{\mu_1=S+1}^{3S} \!\!\!A_{\mu_1}P^{(\bs{\mu_1})}_{i,i+1,i+2} ~+~ \beta
\!\!\sum_{\mu_2=S+\frac{3}{2}}^{2S} \!\!\!A_{\mu_2}P^{(\bs{\mu_2})}_{i,i+1} \right)
\end{equation}
We have restricted $S$ to be half--integer and $\beta>0$.  The
two--fold degenerate ground state of $\mathcal{H}^{S}$ is explicitly
given by $\ket{\psi_{(S\pm\frac{1}{2},S\mp\frac{1}{2})}}$ by means of
\eqref{gs-general}.  For $S=1/2$ the weakly dimerized states reduce to
the well--known Majumdar--Ghosh (MG) states (in this case, the
``weakly dimerized'' states coincide with the completely dimerized
states).  Since the second sum in \eqref{ham-general} does not
contribute for $S=1/2$ the Hamiltonian \eqref{ham-general} reduces to
the sum over the projection operator onto the $S=3/2$ subspace on
three neighboring sites which is known to be the MG Hamiltonian. For
$S=3/2$, the Hamiltonian $\mathcal{H}^{S}$ is equivalent to
\eqref{ham:H1+betaH2}. For larger $S$, \eg $S=5/2$, we find easily the
Hamiltonian $\tilde{\mathcal{H}}_1=\mathcal{H}_1(\frac{5}{2})$ which
has a $2S+1$--fold degenerate ground state manifold consisting of all
VBS states.  Again the term $\tilde{\mathcal{H}}_2$ lifts up all VBS
to higher energy except the weakly dimerized states. For $S=5/2$, the
Hamiltonian $\tilde{\mathcal{H}}_2$ is given by
$\tilde{\mathcal{H}}_2=\sum_i
[\,13397(\bs{S}_i\bs{S}_{i+1})+3582(\bs{S}_i\bs{S}_{i+1})^2
+400(\bs{S}_i\bs{S}_{i+1})^3+16(\bs{S}_i\bs{S}_{i+1})^4+274505/16\,]$.
Although one is restricted to small chain lengths when $S$ becomes
$5/2$ or larger, all these terms are implementable within ED and the
$S=5/2$ spin--dimer model $\tilde{\mathcal{H}}_1+\beta
\tilde{\mathcal{H}}_2$ might be verified numerically.

\section{Conclusion}
In conclusion, we have presented a parent Hamiltonian for the weakly
dimerized VBS states with arbitrary half--integral $S$. For spin 3/2,
we have explained its construction explicitly and have discussed its
main properties. These weakly dimerized spin 3/2 VBS states might be
realized in the impurity induced dimer phase of the ground state of
the molecular nanomagnet Cr8Ni. The path for further numerical
consideration has been pointed out. Finally, the generalization to
spin $S$ has been explained.

\section{Acknowledgement}
I am grateful to Martin Greiter for numerous discussions and his
encouragement for this paper.  I am also indebted to Peter W\"olfle
and Hong-Hao Tu for several helpful suggestions.


\begin{thebibliography}{10}

\bibitem{affleck-87prl799} 
I. Affleck, T. Kennedy, E.~H. Lieb, and H. Tasaki, Phys. Rev. Lett. 
{\bf 59}, 799 (1987).

\bibitem{affleck-88cmp477} 
I. Affleck, T. Kennedy, E.~H. Lieb, and H. Tasaki, 
Commun. Math. Phys. {\bf  115},  477  (1988).

\bibitem{majumdar-69jmp1388}
C.~K. Majumdar and D.~K. Ghosh, J. Math. Phys. {\bf 10}, 1388 (1969).

\bibitem{majumdar-69jmp1399}
C.~K. Majumdar and D.~K. Ghosh, J. Math. Phys. {\bf 10}, 1399 (1969).

\bibitem{affleck89jpcm3047}
I. Affleck, J. Phys.: Condens. Matter {\bf 1},  3047  (1989).

\bibitem{guo-90prb9592}
D. Guo, T. Kennedy, and S. Mazumdar, Phys. Rev. B {\bf 41}, R9592 (1990).

\bibitem{niggemann-96zpb289}
H. Niggemann and J. Zittartz, Z. Phys. B {\bf 101}, 289 (1996).

\bibitem{oshikawa-97prl1984}
M. Oshikawa, M. Yamanaka, and I. Affleck, Phys. Rev. Lett. {\bf 78}, 
1984 (1997).

\bibitem{roth-98prb9264}
R. Roth and U. Schollw\"ock, Phys. Rev. B {\bf 58}, 9264 (1998).

\bibitem{nakamura-02prl077204}
M. Nakamura and S. Todo, Phys. Rev. Lett. {\bf 89}, 077204 (2002).

\bibitem{almeida-08prb094415}
J. Almeida, M.~A. Martin-Delgado, and G. Sierra, Phys. Rev. B {\bf 77}, 
094415 (2008).

\bibitem{almeida-09prb115141}
J. Almeida, M.~A. Martin-Delgado, and G. Sierra, Phys. Rev. B {\bf 79}, 115141 (2009).

\bibitem{refael-02prb060402}
G. Refael, S. Kehrein, and D.~S. Fisher, Phys. Rev. B {\bf 66}, 
060402(R) (2002).

\bibitem{chen-05prb214428}
S. Chen, C. Wu, S.-C. Zhang, and Y. Wang, Phys. Rev. B {\bf 72}, 214428 (2005).

\bibitem{greiter-07prb060401}
M. Greiter, S. Rachel, and D. Schuricht, Phys. Rev. B {\bf 75}, 
060401(R) (2007).

\bibitem{greiter-07prb188841}
M. Greiter and S. Rachel, Phys. Rev. B {\bf 75}, 188841 (2007).

\bibitem{arovas-08prb104404}
D.~P. Arovas, Phys. Rev. B {\bf 77}, 104404 (2008).

\bibitem{tu-08jpa415201}
H.-H. Tu and G.-M. Zhang and T. Xiang, J. Phys. A: Math. Theor. 
{\bf 41} (2008) 415201

\bibitem{tu-08prb094404}
H.-H. Tu and G.-M. Zhang and T. Xiang, Phys. Rev. B {\bf 78}, 
094404 (2008).

\bibitem{schuricht-08prb014430}
D. Schuricht and S. Rachel, Phys. Rev. B {\bf 78}, 014430 (2008).

\bibitem{arovas-09arXiv:0901.1498}
D.~P. Arovas, K. Hasebe, X.-L. Qi, and S.-C. Zhang, preprint 
arXiv:0901.1498 (2009).



\bibitem{schwinger65}
J. Schwinger in L. Biedenharn and H. van Dam (eds), {\it quantum 
Theory of Angular Momentum} (Academic Press, 1965).

\bibitem{auerbach94} 
A. Auerbach, {\em Interacting electrons and quantum magnetism} 
  (Springer, New York, 1994).

\bibitem{arovas-08arXiv:0809.4836}
D.~P. Arovas and A. Auerbach, preprint arXiv:0809.4836 (2008).

\bibitem{arovas-88prl531}
D.~P. Arovas, A. Auerbach, and F.~D.~M. Haldane, 
Phys. Rev. Lett. {\bf 60}, 531 (1988).

\bibitem{qin-95prb12844}
S. Qin, T.-K. Ng, Z.-B. Su, Phys. Rev. B {\bf 52}, 12844 (1995).

\bibitem{hallberg-96prl4955}
K. Hallberg, X.~Q. Wang, P. Horsch, and A. Moreo, Phys. Rev. Lett. 
{\bf 76}, 4955 (1996).

\bibitem{carlon-04prb144416}
E. Carlon, P. Lajk\'o, H. Rieger, F. Igl\'oi, Phys. Rev. B {\bf 69}, 
144416 (2004).

\bibitem{fath-06prb214447}
G. F\'ath, \"O. Legeza, P. Lajk\'o, and F. Igl\'oi, Phys. Rev. B {\bf 73}, 
214447 (2006). 

\bibitem{heidrich-meisner-07prb064413}
F. Heidrich-Meisner, I.~A. Sergienko, A.~E. Feiguin, and E.~R. Dagotto, 
Phys. Rev. B {\bf 75}, 064413 (2007).

\bibitem{sen-07prb104411}
D. Sen and N. Surendran, Phys. Rev. B {\bf 75}, 104411 (2007).

\bibitem{haldane83pla464}
F.~D.~M. Haldane, Phys. Lett.~A {\bf 93},  464  (1983); 
Phys. Rev. Lett. {\bf 50},  1153  (1983).

\bibitem{greiter02jltp1029}
M. Greiter, J. Low Temp. Phys. {\bf 126}, 1029 (2002). 

\bibitem{okamoto-92pla433}
K. Okamoto and K. Nomura, Phys. Lett. A {\bf 169}, 433 (1992).

\bibitem{eggert96prbR9612}
S. Eggert, Phys. Rev. B {\bf 54}, R9612 (1996)

\bibitem{jullien-83baps344}
R. Jullien and F.~D.~M. Haldane, Bull. Am. Phys. Soc. {\bf 28}, 344 (1983).

\end{thebibliography}

\end{document}